\documentclass[11pt,draftcls,peerreview,onecolumn]{IEEEtran}

\usepackage{rrprefsIEEE,empheq}
\newcommand{\smallbox}[1]{\mbox{\small$\displaystyle #1$}}

\begin{document}
%
\title{The analytic computability of the Shannon transform for a large class of random matrix channels}
\author{Raj Rao Nadakuditi\thanks{Department of Mathematics, Massachusetts Institute of Technology, Email: raj@mit.edu, Phone: (857) 891 8303, Fax: (617) 253-4358 }
        }

\markboth{The analytic computability of the Shannon transform for a large class of random matrix channels}{Nadakuditi}
\maketitle

\begin{abstract}
We define a class of ``algebraic'' random matrix channels for which one can generically compute the limiting Shannon transform using numerical techniques and often enumerate the low SNR series expansion coefficients in closed form. We describe this class, the coefficient enumeration techniques and compare theory with simulations.

\end{abstract}


%
\IEEEpeerreviewmaketitle

\section{The Shannon transform}
Consider a  multiple input, multiple output (MIMO) communication system with $N_{r}$ receive antennas and $N_{t}$ transmit antennas where the $N_{r} \times 1$ received vector ${\bf y}$ is modelled as
\begin{equation}\label{eq:mimo model}
{\bf y} = {\bf H}{\bf x} + {\bf z}.
\end{equation}
In (\ref{eq:mimo model}), ${\bf H}$ is an $N_{r} \times N_{t}$ sized random matrix whose $(i,j)$-th entry is the complex valued propagation coefficient between the $i$-th receive antenna and $j$-th transmit antenna. The transmitted signal is denoted by the $N_{t} \times 1$ vector ${\bf x}$ while the $N_r \times 1$ vector ${\bf z}$ is the additive noise at the receiver. We assume that ${\bf z}$ is zero mean, circularly symmetric complex Gaussian noise with independent, equal variance real and imaginary parts and that, without loss of generality, $\mathbb{E}[{\bf x}{\bf x}^{H}]= {\bf I}$ where $\mathbb{E}[.]$ denotes the expectation of the (random) quantity in the brackets. The transmitted vector ${\bf x}$ is subject to the power constraint $P$ so that ${\rm Tr}\,\mathbb{E}[{\bf x}{\bf x}^{H}] \leq P$. 

When the MIMO channel matrix is a random matrix, \ie, its elements are random variables, then it is common to transmit complex valued circularly symmetric signals ${\bf x}$ so that $\mathbb{E}[{\bf x}{\bf x}^{H}] = (P/N_{t}) {\bf I}$. The ergodic capacity of the MIMO system \cite{telatar99a}, assuming the receiver has perfect knowledge of the realization of ${\bf H}$, is then given by 
\begin{equation}\label{eq:mimo capacity}
C(P) := \mathbb{E}_{{\bf H}}\left[\log \det ({\bf I}+ \frac{P}{N_t} {\bf H}{\bf H}{'})\right]
\end{equation}
where $'$ denotes the conjugate transpose and the expectation is with respect to the probability distribution of the random channel matrix ${\bf H}$. Equation (\ref{eq:mimo capacity}) can be rewritten in terms of the eigenvalues of $(1/N_{t}){\bf H}{\bf H}'$ as 
\begin{equation}\label{eq:mimo capacity eig}
C(P) = N_{t}\, \overline{\mathcal{V}}(P) 
\end{equation}
where $\overline{\mathcal{V}}(P)$ is the Shannon transform \cite{tulino04a} of the matrix $(1/N_{t}){\bf H}{\bf H}'$ defined as
\begin{equation}\label{eq:shannon transform}
\overline{\mathcal{V}}(\gamma):= \mathbb{E}_{\lambda}[\log(1+\gamma\,\lambda)],
\end{equation}
and the expectation is with respect to the probability distribution of a randomly selected (with uniform probability) eigenvalue of $(1/N_{t}){\bf H}{\bf H}'$. From (\ref{eq:mimo capacity eig}) and (\ref{eq:shannon transform}) it is evident that one can seek to analytically characterize the Shannon transform of those MIMO random matrix channels for which the eigenvalues of $(1/N_{t}){\bf H}{\bf H}'$ can be analytically characterized. In general, except for the special cases when $N_{t} = 1$ or $N_{r} = 1$, the ``exact'' analytical expressions for Shannon transform found in the literature \cite{telatar99a,martin04a,moewin04a,moewin06a} are really determinental formulae. In other words, the Shannon transform is expressed as a determinant of a matrix for which there are closed form expressions for the individual elements. The reader is directed to \cite{alfano06a} for some representative formulae and a summary of random matrix channels for which exact closed form expressions are available. 

The existence of the determinental representation for the Shannon transform is a fundamental truism even in the simplest case when there is i.i.d. Rayleigh fading between the transmitter and receiver antenna elements \cite{telatar99a}. A limitation of these results is that the determinental end results makes it is hard for practitioners to gain engineering insight on how the parameters such as $SNR$, $N_{r}$ and $N_{t}$  affect the ergodic capacity (Shannon transform). More importantly, the range of random matrix channels for which the eigenvalues of $(1/N_{t}){\bf H}{\bf H}'$ can be characterized exactly for \textit{finite} $N_{r}, N_{t}$ is very restrictive and invariably limited to matrices with (complex) Gaussian entries. 

\subsection{The limiting Shannon transform}
This has motivated the investigation into the properties of the limiting Shannon transform $\mathcal{V}(\gamma)$ instead which is defined as
\begin{equation}
\mathcal{V}(\gamma) \equiv \mathcal{V}_{\infty}(\gamma) := \lim_{N_{r}, N_{t} \to \infty} \overline{\mathcal{V}}(\gamma) \qquad \textrm{for} \qquad N_{r}/N_{t} \to c \in (0,\infty).
\end{equation}
Let the empirical distribution function (e.d.f.) of an arbitrary $N \times N$  matrix ${\bf A}_{N}$ with real eigenvalues be defined as
\begin{equation}\label{eq:intro edf}
F^{{\bf A}_{N}}(x)=\frac{\textrm{Number of eigenvalues of }{\bf A}_{N} \leq x }{N}.
\end{equation}
If the (random) e.d.f of ${\bf W} := (1/N_{t}){\bf H}{\bf H}'$ converges, for every $x$, almost surely (or in probability) as $N_{t},N_{r}(N_{t}) \to \infty$ to a non-random distribution function $F^{W}(x)$, then the limiting Shannon transform, when the limit exists, can be written as
\begin{equation}\label{eq:limiting St}
\mathcal{V}_{W}(\gamma) = \int \log(1+\gamma\lambda) dF^{W}(\lambda).
\end{equation}
The limiting Shannon transform, for small values of $\gamma$, can be expressed as the series
\begin{align}\label{eq:St series}
\mathcal{V}_{W}(\gamma) &= \int \sum_{k=1}^{\infty} \frac{(-1)^{k+1}}{k} \gamma^{k} \lambda^{k} dF^{W}(\lambda) = \sum_{k=1}^{\infty}  \underbrace{\frac{(-1)^{k+1}}{k} M^{W}_{k}}_{\nu^{W}_{k}} \gamma^{k} 
\end{align}
where $M^{W}_{k}:=\int \lambda^{k} dF^{W}(\lambda)$ is the $k$-th moment of the limiting probability distribution function $F^{W}$. 

The main contribution of this correspondence, which relies on the results in \cite{raj:thesis,raj04a}, is the identification of a broad class of random matrix channels for which the limiting Shannon transform in (\ref{eq:limiting St}) can be numerically computed and for which the coefficients of the series expansion in (\ref{eq:St series}) can be efficiently enumerated, often in closed form. Examples found in the literature \cite{telatar99a,muller02a,muller02b,mestre03a,skupch05a} which rely on results from infinite/large random matrix theory are but special cases of this broader class of algebraic random matrix channels, which we define next. We leave it to practitioners to justify the physical relevance of more complicated random matrix models built using the framework presented.

\section{The class of algebraic MIMO channels}

\begin{definitionS}[Algebraic random matrix \cite{raj:thesis,raj04a}]
Let $F^{W}(x)$ denote the limiting eigenvalue distribution function of a sequence of random matrices ${\bf W}_{N}$.  If a bivariate polynomial $$L_{{\rm mz}}(m,z)=\sum_{i=0}^{D_{{\rm m}}} \sum_{j=0}^{D_{{\rm z}}} a_{ij} m^{i} z^{j},$$ with $D_{{\rm m}}>0$, $D_{{\rm z}}>0$ and real-valued coefficients $a_{ij}$ exists such that the Stieltjes transform of $F^{W}(x)$ defined as
\begin{equation}\label{eq:mz}
m_{W}(z) = \int \frac{1}{x-z} dF^{W}(x) \qquad \textrm{for} \qquad z \in \mathbb{C}^{+} \setminus \mathbb{R},
\end{equation}
is algebraic, \ie, it is a solution of the equation $L_{{\rm mz}}(m_{W}(z),z) = 0$ then ${\bf W}_{N}$ is said to be an algebraic random matrix. The density function $f_{W}=dF^{W}$ is referred to as an algebraic density and we say that $f_{W} \in \mathcal{P}_{{\rm alg}}$ and ${\bf W}_{N} \in \mathcal{M}_{{\rm alg}}$, the class of algebraic probability densities and random matrices respectively.\\
\end{definitionS}

\begin{definitionS}[Algebraic MIMO random matrix channel]\label{def:intro Malg}
Let ${\bf H}$ be an $N_{r} \times N_{t}$ sized random matrix MIMO channel. If ${\bf W}_{N_{t}}:=(1/N_{t}){\bf H}{\bf H}'$ is an algebraic random matrix  for $N_{t},N_{r}(N_{t}) \infty$ and $N_{r}/N_{t} \to c>0$ then ${\bf H}$ is said to be an algebraic MIMO random matrix channel and we say that ${\bf H} \in \mathcal{H}_{{\rm alg}}$, the class of algebraic  MIMO channels.\\
\end{definitionS}

In \cite{raj:thesis,raj06a}, we describe the generators of the class of algebraic random matrices as well as procedures for computing the bivariate polynomial  $\lmzs{W}$ that encodes the limiting eigenvalue distribution. We focus on the sub-class of random matrix channels that are generated from Gaussian distributed entries in what follows. We direct the reader to \cite{raj:thesis,raj06a} for additional examples.

\theorembox{
\begin{theorem}[\cite{raj:thesis,raj04a}]\label{th:main th}
Assume that ${\bf G}$ is an $N_{r} \times N_{t}$ sized MIMO random matrix channel with i.i.d. $\mathcal{CN}(0,1)$ distributed elements and that ${\bf A}$ and ${\bf B}$ are appropriately sized non-negative definite algebraic random matrices independent of ${\bf G}$. Then for all $s > 0$, the MIMO random matrix channels
\begin{itemize}
\item Doubly correlated channel model: ${\bf H} = {\bf A}^{1/2} {\bf G} {\bf B}^{1/2}$
\item Random Rician-like fading model: ${\bf H} =  {\bf A}^{1/2} + s\,{\bf G}$ 
\end{itemize}
are algebraic as well in the sense of Definition \ref{def:intro Malg}.
\end{theorem}
}

Theorem \ref{th:main th} provides the building block for analyzing the properties of a much broader classes of random matrices than what is found in the literature. The starting point for applying Theorem \ref{th:main th} is the bivariate polynomial representations $\lmzs{A}$ and $\lmzs{B}$. In \cite{raj:thesis,raj06a} we describe the mapping $\lmzs{A},\lmzs{B} \mapsto \lmzs{W}$ and implement it in the form of a \mat based random matrix calculator which can be downloaded from \cite{raj:rmtool}. In particular, the commands \texttt{corrWish(LmzA,LmzB,c)} and \texttt{AgramWish(LmzA,c,s)} implement the mapping $\lmzs{A},\lmzs{B} \mapsto \lmzs{W}$ for the random matrix transformations in Theorem \ref{th:main th}.

To make the ideas presented more concrete we consider some simple random matrix channels and list the corresponding bivariate polynomial encoding. For the channel ${\bf H} = N_{t}$, ${\bf W} = (1/N_{t}){\bf H}{\bf H}' = {\bf I}$ so that the Stieltjes transform of $F^{W}(x) = \mathbb{I}_{(1,\infty)}(x)$, defined as in (\ref{eq:mz}), can be shown to satisfy the equation $\lmz{W} = 0$ where
\[
\lmz{w} = m(1-z) - 1.
\]
The Rayleigh fading channel considered in \cite{telatar99a} is algebraic by applying Theorem \ref{th:main th} with ${\bf A} = {\bf I}_{N_{r}}$ and  ${\bf B} = {\bf I}_{N_{t}}$. It can be shown that 

\begin{equation}\label{eq:ex1}
\lmz{W} = cz{m}^{2}- \left( 1-c-z \right) m +1
\end{equation}

The doubly correlated Gaussian random matrix falls into the setting described in Theorem \ref{th:main th}. The situation where matrices ${\bf A}$ and ${\bf B}$ are the covariances of an AR(1) process with coefficient $\alpha$ is considered in \cite{skupch05a} and is yet another example of an algebraic random matrix channel. Here we have 
\[
\lmz{A} = \lmz{B} = \left({z}^{3}-2\,{z}^{2}\alpha+z \right) {m}^{2}+ \left( 2\,{z}^{2}-4\,\alpha\,z+2 \right) m+z-2\,\alpha
\]
and 
\begin{multline*}
\lmz{W}=-{z}^{3}{m}^{4}{c}^{2}+ \left( 2\,{z}^{2}c-2\,{z}^{3}\alpha\,c-4\,{z}^{2}{c}^{2} \right) {m}^{3}+ \left( 2\,{z}^{2}\alpha-{z}^{3}-z-5\,z{c}^{2}-6\,{z}^{2}\alpha\,c+6\,cz \right) {m}^{2} \\
+ \left( -6\,\alpha\,zc+4\,\alpha\,z-2-2\,{z}^{2}-2\,{c}^{2}+4\,c \right) m-2\,\alpha\,c-z+2\,\alpha
\end{multline*}

Consider the situation when the matrices ${\bf A}$ and ${\bf B}$ have limiting eigenvalue distribution
\begin{equation}\label{eq:Fa ex}
F^{A}(x) = F^{B}(x) = 0.5 \mathbb{I}_{[1,\infty)} + 0.5 \mathbb{I}_{[2,\infty)}
\end{equation}
so that their Stieltjes transform satisfies the equation $\lmz{A} = 0 = \lmz{B}$ where
\[
\lmz{A} = \lmz{B} =  \left( -6\,z+2\,{z}^{2}+4 \right) m+2\,z-3.
\]
Then the random matrix channel ${\bf A}^{1/2}G{\bf B}^{1/2}$ is algebraic and we have 

\begin{equation}\label{eq:ex2}
\lmzs{W} = \sum_{j=1}^{6}\sum_{k=1}^{4} \left[{\bf T}_{{\rm mz}}^{C} \right]_{{\rm jk}} m^{j-1} z^{k-1},
\end{equation}
where:
\begin{equation}
\smallbox{{\bf T}_{{\rm mz}}^{C} \equiv 
 \left[ \begin {array}{cccc} -18\,c+18\,{c}^{2}&18\,c-9&4&0\\\noalign{\medskip}-108\,{c}^{2}+36\,c+72\,{c}^{3}&-112\,c+18+130\,{c}^{2}&-18+54\,c&4\\\noalign{\medskip}64\,{c}^{2}+64\,{c}^{4}-128\,{c}^{3}&72\,c-324\,{c}^{2}+288\,{c}^{3}&224\,{c}^{2}-112\,c&36\,c\\\noalign{\medskip}0&64\,{c}^{2}-256\,{c}^{3}+192\,{c}^{4}&360\,{c}^{3}-216\,{c}^{2}&112\,{c}^{2}\\\noalign{\medskip}0&0&192\,{c}^{4}-128\,{c}^{3}&144\,{c}^{3}\\\noalign{\medskip}0&0&0&64\,{c}^{4}\end {array} \right].
}\\
\end{equation}
We leave it to the reader to verify that the examples considered in  \cite{muller02a,muller02b,mestre03a} are special cases of algebraic random matrix channels as well.

\subsection{Computation of the Shannon transform and its low SNR series expansion}

Once we have $\lmzs{W}$, we can obtain the limiting eigenvalue distribution by a simple root-finding algorithm as described in \cite{raj06a,raj:thesis}, isolating the correct branch of the $\Dm$ solutions, taking its imaginary part and scaling by $1/\pi$. This is motivated by the fact that the probability distribution $F^{W}$ can be recovered from its Stieltjes transform by using the Stieltjes inversion formula \cite{akhiezer65a}   
\begin{equation}\label{eq:inversion formula}
dF^{W}(x) =\frac{1}{\pi} \lim _{\xi \rightarrow 0^{+}}{\rm Im} \:m_{W}(x+i\xi).
\end{equation}
In the examples considered above, except for the simplest case corresponding to i.i.d. Rayleigh fading, $\Dm \geq 4$ so that we have to resort to numerical techniques to obtain $dF^{W}(x)$ and compute the Shannon transform using (\ref{eq:limiting St}). The development of efficient numerical code that extracts the correct branch of the Stieltjes transform from the $\Dm$ solutions of the equation $\lmz{W}=0$ so that (\ref{eq:inversion formula}) may be applied to yield the limiting eigenalue distribution remains an open problem. Hence, the remarkable fact \cite{raj06a,raj:thesis} that for algebraic random matrix channels it will generically be possible to obtain the coefficients of the low SNR series expansion in (\ref{eq:St series}) in closed form assumes greater importance as far as lending engineering insight when dealing with complicated (algebraic) MIMO channel models. 

We note that $\nu^{W}_{k}$ the $k$-th coefficient of the Shannon transform series expansion in (\ref{eq:St series}) is given by $\nu^{W}_{k} = (-1)^{k+1}M^{W}_{k}/k$. The algebraicity of the limiting eigenvalue distribution allows us to efficiently enumerate the limiting moments in closed form. To do so, we first define the ordinary) moment generating function $\mu_{W}(z) := 1+\sum_{i=1}^{\infty} M_{k}^{W}$ and note that it can be obtained from the Stieltjes transform $m_{W}(z)$ by applying the transformation
\begin{equation}
\mu_{W}(z) = \frac{1}{z}m_{W}(1/z).
\end{equation}
Thus, given the bivariate polynomial $\lmz{W}$ we can obtain the algebraic equation $\lmuz{W}$ satisfied by $\mu_{W}(z)$ by applying the transformation
$$\lmuz{}= \lmzs{}(-\mu z, 1/z)$$
and clearing the denominator. The \maple based package \texttt{gfun} \cite{salvy94a} can be used to obtain the series expansion for $\mu_{W}(z)$ up to degree \texttt{expansion\_degree} directly from the bivariate polynomial $\lmuzs{}$ by using the commands:

\begin{small}
\begin{verbatim}
   > with(gfun):
   > MomentSeries = algeqtoseries(Lmyuz,z,myu,expansion_degree,'pos_slopes');
\end{verbatim}
\end{small}

For the i.i.d. Rayleigh fading channel whose limiting eigenvalue distribution is encoded by the bivariate polynomial (\ref{eq:ex1}), the corresponding moment generating series is given by 
\begin{equation}
\mu_{W} = 1 + z + (1+c) z^{2} + (1+ 3c +c^{2}) z^{3} + (1+6c + 6c^2 + c^{3}) z^{4} + O(z^{5})
\end{equation}
For the doubly correlated Rayleigh fading channel whose limiting eigenvalue distribution is encoded by the bivariate polynomial (\ref{eq:ex2}), the corresponding moment generating series is given by 
\begin{multline}
\mu_{W}(z)=1+{\frac {9}{4}}z+ \left( {\frac {45}{8}}\,c+{\frac {45}{8}} \right) {z}^{2}+ \left( {\frac {675}{16}}\,c+{\frac {243}{
16}}\,{c}^{2}+{\frac {243}{16}} \right) {z}^{3}\\
+ \left( {\frac {3555}{16}}\,{c}^{2}+{\frac {1377}{32}}\,{c}^{3}+{\frac {
3555}{16}}\,c+{\frac {1377}{32}} \right) {z}^{4}+O \left( {z}^{5} \right). 
\end{multline}

\section{Numerical simulations}

\begin{figure}
\centering
\includegraphics[width=5.5in]{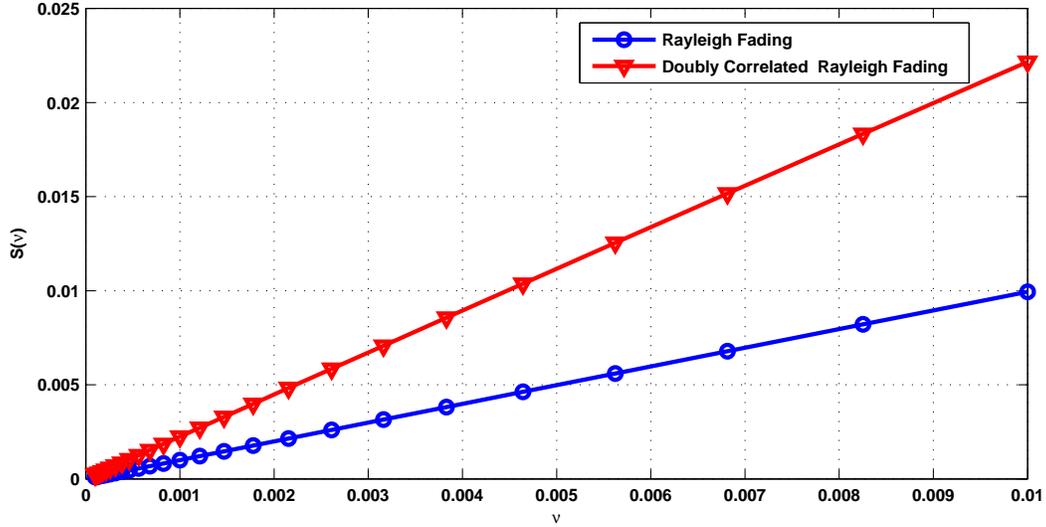}
\caption{The mean Shannon transform versus $\gamma$ averaged over 20000 trials for $N_{r}=50, N_{t} = 200$.}
\label{fig:numerics}
\end{figure}

Figure \ref{fig:numerics} plots the mean empirical Shannon transform $\mathcal{V}(\gamma)$ for various values of $\gamma$ (SNR) for the i.i.d. (uncorrelated) Rayleigh fading channel and the doubly correlated Rayleigh fading channel with the limiting eigenvalue distributions of ${\bf A}$ and ${\bf B}$ given by (\ref{eq:Fa ex}). The Table \ref{tab:rayleigh2 numerics} compares the coefficients of the series expansion obtained from the empirical data with the theoretical predictions. The excellent agreement confirms the utility of the closed form expansions and the well document fact \cite{tulino04a}that the $N_{r}, N_{t} \to \infty$ limiting answer is a good approximation of the $N_r, N_t$ finite result.

\begin{table}
\centering
\subtable[I.i.d. Rayleigh fading.]{
\label{tab:rayleigh numerics}
\begin{tabular}{|c|c|c||c|c||c|c||c|c|}
\hline
$N_{r}$& $N_{t}$  &c &$\widehat{\nu}_{1}$ &$\nu_{1}=1$      &$\widehat{\nu}_{2}$     & $\nu_{2}=-(1+c)/2$      &$\widehat{\nu}_{3}$  &$\nu_{3}=(1+3c+c^2)/3$  \\ 
\hline
50 & 200 &0.25  &1.0000  &1.0000  &-0.6250 &-0.6250 &0.5989   &0.6042 \\
50 & 100 &0.50  &1.0001  &1.0000  &-0.7502 &-0.7500 &0.9070   &0.9167 \\
50  & 50 &1     &1.0003  &1.0000  &-1.0004 &-1.0000 &1.6430   &1.6667 \\
50 & 26  &1.923 &0.9998  &1.0000  &-1.4609 &-1.4615 &3.4145   &3.4892 \\
\hline
\end{tabular}}\\
\subtable[Doubly correlated Rayleigh fading where the matrices ${\bf A}$ and ${\bf B}$ have limiting e.d.f given by (\ref{eq:Fa ex}).]{
\label{tab:rayleigh2 numerics}
\begin{tabular}{|c|c|c||c|c||c|c||c|c|}
\hline
$N_{r}$& $N_{t}$  &c &$\widehat{\nu}_{1}$ &$\nu_{1}=9/4$      &$\widehat{\nu}_{2}$     & $\nu_{2}=-\left( {\frac {45}{16}}\,c+{\frac {45}{16}}\right)$      &$\widehat{\nu}_{3}$  &$\nu_{3}=\left( {\frac {675}{48}}\,c+{\frac {243}{
48}}\,{c}^{2}+{\frac {243}{48}} \right)$  \\
\hline
50 & 200 &0.25 &2.2500  &2.2500  &-3.5153 &-3.5156  &8.6956   &8.8945 \\
50 & 100 &0.5  &2.2502  &2.2500  &-4.2189 &-4.2188  &12.9866  &13.3594 \\
50  & 50 &1    &2.2509  &2.2500  &-5.6276 &-5.6250  &23.2916  &24.1875 \\
50 & 26  &1.923 &2.2494  &2.2500  &-8.2139 &-8.2212  &48.0846  &50.8280 \\
\hline
\end{tabular}}
\caption{Comparison of theoretical $\nu_{k}$ in (\ref{eq:St series}) with estimates from numerical simulations}
\end{table}

\section*{Acknowledgements}
The author's work was supported by an ONR Postdoctoral Special Research Award in Ocean Acoustics under grant N00014-07-1-0269. The author thanks Arthur Baggeroer for his encouragement.

%
%



\bibliographystyle{IEEEtran}
\bibliography{randbib}
%
\end{document}